\documentclass[10pt,conference,twocolumn]{IEEEtran}

\usepackage{cite}
\usepackage[T1]{fontenc}
\usepackage{graphicx}
\usepackage{amssymb}
\usepackage{amsmath}
\usepackage{amsthm}
 
\usepackage{microtype}
\usepackage{balance}
\usepackage{subfigure}
\usepackage{xcolor}
\usepackage{booktabs}
\usepackage{algorithm}
\usepackage[noend]{algpseudocode}

\usepackage{amssymb}
\usepackage{amsfonts}
\usepackage{mathrsfs}
\usepackage{xspace}
\usepackage{bm}
\usepackage{upgreek}

\newcommand{\safemath}[2]{\newcommand{#1}{\ensuremath{#2}\xspace}}

\safemath{\bma}{\mathbf{a}}
\safemath{\bmb}{\mathbf{b}}
\safemath{\bmc}{\mathbf{c}}
\safemath{\bmd}{\mathbf{d}}
\safemath{\bme}{\mathbf{e}}
\safemath{\bmf}{\mathbf{f}}
\safemath{\bmg}{\mathbf{g}}
\safemath{\bmh}{\mathbf{h}}
\safemath{\bmi}{\mathbf{i}}
\safemath{\bmj}{\mathbf{j}}
\safemath{\bmk}{\mathbf{k}}
\safemath{\bml}{\mathbf{l}}
\safemath{\bmm}{\mathbf{m}}
\safemath{\bmn}{\mathbf{n}}
\safemath{\bmo}{\mathbf{o}}
\safemath{\bmp}{\mathbf{p}}
\safemath{\bmq}{\mathbf{q}}
\safemath{\bmr}{\mathbf{r}}
\safemath{\bms}{\mathbf{s}}
\safemath{\bmt}{\mathbf{t}}
\safemath{\bmu}{\mathbf{u}}
\safemath{\bmv}{\mathbf{v}}
\safemath{\bmw}{\mathbf{w}}
\safemath{\bmx}{\mathbf{x}}
\safemath{\bmy}{\mathbf{y}}
\safemath{\bmz}{\mathbf{z}}
\safemath{\bmzero}{\mathbf{0}}
\safemath{\bmone}{\mathbf{1}}

\bmdefine{\biad}{a}
\bmdefine{\bibd}{b}
\bmdefine{\bicd}{c}
\bmdefine{\bidd}{d}
\bmdefine{\bied}{e}
\bmdefine{\bifd}{f}
\bmdefine{\bigd}{g}
\bmdefine{\bihd}{h}
\bmdefine{\biid}{i}
\bmdefine{\bijd}{j}
\bmdefine{\bikd}{k}
\bmdefine{\bild}{l}
\bmdefine{\bimd}{m}
\bmdefine{\bind}{n}
\bmdefine{\biod}{o}
\bmdefine{\bipd}{p}
\bmdefine{\biqd}{q}
\bmdefine{\bird}{r}
\bmdefine{\bisd}{s}
\bmdefine{\bitd}{t}
\bmdefine{\biud}{u}
\bmdefine{\bivd}{v}
\bmdefine{\biwd}{w}
\bmdefine{\bixd}{x}
\bmdefine{\biyd}{y}
\bmdefine{\bizd}{z}

\bmdefine{\bixid}{\xi}
\bmdefine{\bilambdad}{\lambda}
\bmdefine{\bimud}{\mu}
\bmdefine{\bithetad}{\theta}
\bmdefine{\biphid}{\phi}
\bmdefine{\bideltad}{\delta}

\safemath{\bmia}{\biad}
\safemath{\bmib}{\bibd}
\safemath{\bmic}{\bicd}
\safemath{\bmid}{\bidd}
\safemath{\bmie}{\bied}
\safemath{\bmif}{\bifd}
\safemath{\bmig}{\bigd}
\safemath{\bmih}{\bihd}
\safemath{\bmii}{\biid}
\safemath{\bmij}{\bijd}
\safemath{\bmik}{\bikd}
\safemath{\bmil}{\bild}
\safemath{\bmim}{\bimd}
\safemath{\bmin}{\bind}
\safemath{\bmio}{\biod}
\safemath{\bmip}{\bipd}
\safemath{\bmiq}{\biqd}
\safemath{\bmir}{\bird}
\safemath{\bmis}{\bisd}
\safemath{\bmit}{\bitd}
\safemath{\bmiu}{\biud}
\safemath{\bmiv}{\bivd}
\safemath{\bmiw}{\biwd}
\safemath{\bmix}{\bixd}
\safemath{\bmiy}{\biyd}
\safemath{\bmiz}{\bizd}

\safemath{\bmxi}{\bixid}
\safemath{\bmlambda}{\bilambdad}
\safemath{\bmmu}{\bimud}
\safemath{\bmtheta}{\bithetad}
\safemath{\bmphi}{\biphid}
\safemath{\bmdelta}{\bideltad}

\safemath{\bA}{\mathbf{A}}
\safemath{\bB}{\mathbf{B}}
\safemath{\bC}{\mathbf{C}}
\safemath{\bD}{\mathbf{D}}
\safemath{\bE}{\mathbf{E}}
\safemath{\bF}{\mathbf{F}}
\safemath{\bG}{\mathbf{G}}
\safemath{\bH}{\mathbf{H}}
\safemath{\bI}{\mathbf{I}}
\safemath{\bJ}{\mathbf{J}}
\safemath{\bK}{\mathbf{K}}
\safemath{\bL}{\mathbf{L}}
\safemath{\bM}{\mathbf{M}}
\safemath{\bN}{\mathbf{N}}
\safemath{\bO}{\mathbf{O}}
\safemath{\bP}{\mathbf{P}}
\safemath{\bQ}{\mathbf{Q}}
\safemath{\bR}{\mathbf{R}}
\safemath{\bS}{\mathbf{S}}
\safemath{\bT}{\mathbf{T}}
\safemath{\bU}{\mathbf{U}}
\safemath{\bV}{\mathbf{V}}
\safemath{\bW}{\mathbf{W}}
\safemath{\bX}{\mathbf{X}}
\safemath{\bY}{\mathbf{Y}}
\safemath{\bZ}{\mathbf{Z}}

\safemath{\bZero}{\mathbf{0}}
\safemath{\bOne}{\mathbf{1}}
\safemath{\bDelta}{\mathbf{\Delta}}
\safemath{\bLambda}{\mathbf{\UpLambda}}
\safemath{\bPhi}{\mathbf{\Upphi}}
\safemath{\bSigma}{\mathbf{\Upsigma}}
\safemath{\bOmega}{\mathbf{\Upomega}}
\safemath{\bTheta}{\mathbf{\Uptheta}}

\bmdefine{\biAd}{A}
\bmdefine{\biBd}{B}
\bmdefine{\biCd}{C}
\bmdefine{\biDd}{D}
\bmdefine{\biEd}{E}
\bmdefine{\biFd}{F}
\bmdefine{\biGd}{G}
\bmdefine{\biHd}{H}
\bmdefine{\biId}{I}
\bmdefine{\biJd}{J}
\bmdefine{\biKd}{K}
\bmdefine{\biLd}{L}
\bmdefine{\biMd}{M}
\bmdefine{\biOd}{N}
\bmdefine{\biPd}{O}
\bmdefine{\biQd}{P}
\bmdefine{\biRd}{R}
\bmdefine{\biSd}{S}
\bmdefine{\biTd}{T}
\bmdefine{\biUd}{U}
\bmdefine{\biVd}{V}
\bmdefine{\biWd}{W}
\bmdefine{\biXd}{X}
\bmdefine{\biYd}{Y}
\bmdefine{\biZd}{Z}

\bmdefine{\biDelta}{\Delta}
\bmdefine{\biLambda}{\Lambda}
\bmdefine{\biPhi}{\Phi}
\bmdefine{\biSigma}{\Sigma}
\bmdefine{\biOmega}{\Omega}
\bmdefine{\biTheta}{\Theta}

\safemath{\bimA}{\biAd}
\safemath{\bimB}{\biBd}
\safemath{\bimC}{\biCd}
\safemath{\bimD}{\biDd}
\safemath{\bimE}{\biEd}
\safemath{\bimF}{\biFd}
\safemath{\bimG}{\biGd}
\safemath{\bimH}{\biHd}
\safemath{\bimI}{\biId}
\safemath{\bimJ}{\biJd}
\safemath{\bimK}{\biKd}
\safemath{\bimL}{\biLd}
\safemath{\bimM}{\biMd}
\safemath{\bimN}{\biNd}
\safemath{\bimO}{\biOd}
\safemath{\bimP}{\biPd}
\safemath{\bimQ}{\biQd}
\safemath{\bimR}{\biRd}
\safemath{\bimS}{\biSd}
\safemath{\bimT}{\biTd}
\safemath{\bimU}{\biUd}
\safemath{\bimV}{\biVd}
\safemath{\bimW}{\biWd}
\safemath{\bimX}{\biXd}
\safemath{\bimY}{\biYd}
\safemath{\bimZ}{\biZd}

\safemath{\bimDelta}{\biDelta}
\safemath{\bimLambda}{\biLambda}
\safemath{\bimPhi}{\biPhi}
\safemath{\bimSigma}{\biSigma}
\safemath{\bimOmega}{\biOmega}
\safemath{\bimTheta}{\biTheta}

\safemath{\setA}{\mathcal{A}}
\safemath{\setB}{\mathcal{B}}
\safemath{\setC}{\mathcal{C}}
\safemath{\setD}{\mathcal{D}}
\safemath{\setE}{\mathcal{E}}
\safemath{\setF}{\mathcal{F}}
\safemath{\setG}{\mathcal{G}}
\safemath{\setH}{\mathcal{H}}
\safemath{\setI}{\mathcal{I}}
\safemath{\setJ}{\mathcal{J}}
\safemath{\setK}{\mathcal{K}}
\safemath{\setL}{\mathcal{L}}
\safemath{\setM}{\mathcal{M}}
\safemath{\setN}{\mathcal{N}}
\safemath{\setO}{\mathcal{O}}
\safemath{\setP}{\mathcal{P}}
\safemath{\setQ}{\mathcal{Q}}
\safemath{\setR}{\mathcal{R}}
\safemath{\setS}{\mathcal{S}}
\safemath{\setT}{\mathcal{T}}
\safemath{\setU}{\mathcal{U}}
\safemath{\setV}{\mathcal{V}}
\safemath{\setW}{\mathcal{W}}
\safemath{\setX}{\mathcal{X}}
\safemath{\setY}{\mathcal{Y}}
\safemath{\setZ}{\mathcal{Z}}
\safemath{\emptySet}{\varnothing}

\safemath{\colA}{\mathscr{A}}
\safemath{\colB}{\mathscr{B}}
\safemath{\colC}{\mathscr{C}}
\safemath{\colD}{\mathscr{D}}
\safemath{\colE}{\mathscr{E}}
\safemath{\colF}{\mathscr{F}}
\safemath{\colG}{\mathscr{G}}
\safemath{\colH}{\mathscr{H}}
\safemath{\colI}{\mathscr{I}}
\safemath{\colJ}{\mathscr{J}}
\safemath{\colK}{\mathscr{K}}
\safemath{\colL}{\mathscr{L}}
\safemath{\colM}{\mathscr{M}}
\safemath{\colN}{\mathscr{N}}
\safemath{\colO}{\mathscr{O}}
\safemath{\colP}{\mathscr{P}}
\safemath{\colQ}{\mathscr{Q}}
\safemath{\colR}{\mathscr{R}}
\safemath{\colS}{\mathscr{S}}
\safemath{\colT}{\mathscr{T}}
\safemath{\colU}{\mathscr{U}}
\safemath{\colV}{\mathscr{V}}
\safemath{\colW}{\mathscr{W}}
\safemath{\colX}{\mathscr{X}}
\safemath{\colY}{\mathscr{Y}}
\safemath{\colZ}{\mathscr{Z}}

\safemath{\opA}{\mathbb{A}}
\safemath{\opB}{\mathbb{B}}
\safemath{\opC}{\mathbb{C}}
\safemath{\opD}{\mathbb{D}}
\safemath{\opE}{\mathbb{E}}
\safemath{\opF}{\mathbb{F}}
\safemath{\opG}{\mathbb{G}}
\safemath{\opH}{\mathbb{H}}
\safemath{\opI}{\mathbb{I}}
\safemath{\opJ}{\mathbb{J}}
\safemath{\opK}{\mathbb{K}}
\safemath{\opL}{\mathbb{L}}
\safemath{\opM}{\mathbb{M}}
\safemath{\opN}{\mathbb{N}}
\safemath{\opO}{\mathbb{O}}
\safemath{\opP}{\mathbb{P}}
\safemath{\opQ}{\mathbb{Q}}
\safemath{\opR}{\mathbb{R}}
\safemath{\opS}{\mathbb{S}}
\safemath{\opT}{\mathbb{T}}
\safemath{\opU}{\mathbb{U}}
\safemath{\opV}{\mathbb{V}}
\safemath{\opW}{\mathbb{W}}
\safemath{\opX}{\mathbb{X}}
\safemath{\opY}{\mathbb{Y}}
\safemath{\opZ}{\mathbb{Z}}
\safemath{\opZero}{\mathbb{O}}
\safemath{\identityop}{\opI}

\safemath{\veca}{\bma}
\safemath{\vecb}{\bmb}
\safemath{\vecc}{\bmc}
\safemath{\vecd}{\bmd}
\safemath{\vece}{\bme}
\safemath{\vecf}{\bmf}
\safemath{\vecg}{\bmg}
\safemath{\vech}{\bmh}
\safemath{\veci}{\bmi}
\safemath{\vecj}{\bmj}
\safemath{\veck}{\bmk}
\safemath{\vecl}{\bml}
\safemath{\vecm}{\bmm}
\safemath{\vecn}{\bmn}
\safemath{\veco}{\bmo}
\safemath{\vecp}{\bmp}
\safemath{\vecq}{\bmq}
\safemath{\vecr}{\bmr}
\safemath{\vecs}{\bms}
\safemath{\vect}{\bmt}
\safemath{\vecu}{\bmu}
\safemath{\vecv}{\bmv}
\safemath{\vecw}{\bmw}
\safemath{\vecx}{\bmx}
\safemath{\vecy}{\bmy}
\safemath{\vecz}{\bmz}

\safemath{\veczero}{\bmzero}
\safemath{\vecone}{\bmone}
\safemath{\vecxi}{\bmxi}
\safemath{\veclambda}{\bmlambda}
\safemath{\vecmu}{\bmmu}
\safemath{\vectheta}{\bmtheta}
\safemath{\vecphi}{\bmphi}
\safemath{\vecdelta}{\bmdelta}

\safemath{\matA}{\bA}
\safemath{\matB}{\bB}
\safemath{\matC}{\bC}
\safemath{\matD}{\bD}
\safemath{\matE}{\bE}
\safemath{\matF}{\bF}
\safemath{\matG}{\bG}
\safemath{\matH}{\bH}
\safemath{\matI}{\bI}
\safemath{\matJ}{\bJ}
\safemath{\matK}{\bK}
\safemath{\matL}{\bL}
\safemath{\matM}{\bM}
\safemath{\matN}{\bN}
\safemath{\matO}{\bO}
\safemath{\matP}{\bP}
\safemath{\matQ}{\bQ}
\safemath{\matR}{\bR}
\safemath{\matS}{\bS}
\safemath{\matT}{\bT}
\safemath{\matU}{\bU}
\safemath{\matV}{\bV}
\safemath{\matW}{\bW}
\safemath{\matX}{\bX}
\safemath{\matY}{\bY}
\safemath{\matZ}{\bZ}
\safemath{\matzero}{\bmzero}

\safemath{\matDelta}{\bDelta}
\safemath{\matLambda}{\bLambda}
\safemath{\matPhi}{\bPhi}
\safemath{\matSigma}{\bSigma}
\safemath{\matOmega}{\bOmega}
\safemath{\matTheta}{\bTheta}

\safemath{\matidentity}{\matI}
\safemath{\matone}{\matO}

\safemath{\rnda}{A}
\safemath{\rndb}{B}
\safemath{\rndc}{C}
\safemath{\rndd}{D}
\safemath{\rnde}{E}
\safemath{\rndf}{F}
\safemath{\rndg}{G}
\safemath{\rndh}{H}
\safemath{\rndi}{I}
\safemath{\rndj}{J}
\safemath{\rndk}{K}
\safemath{\rndl}{L}
\safemath{\rndm}{M}
\safemath{\rndn}{N}
\safemath{\rndo}{O}
\safemath{\rndp}{P}
\safemath{\rndq}{Q}
\safemath{\rndr}{R}
\safemath{\rnds}{S}
\safemath{\rndt}{T}
\safemath{\rndu}{U}
\safemath{\rndv}{V}
\safemath{\rndw}{W}
\safemath{\rndx}{X}
\safemath{\rndy}{Y}
\safemath{\rndz}{Z}

\safemath{\rveca}{\bimA}
\safemath{\rvecb}{\bimB}
\safemath{\rvecc}{\bimC}
\safemath{\rvecd}{\bimD}
\safemath{\rvece}{\bimE}
\safemath{\rvecf}{\bimF}
\safemath{\rvecg}{\bimG}
\safemath{\rvech}{\bimH}
\safemath{\rveci}{\bimI}
\safemath{\rvecj}{\bimJ}
\safemath{\rveck}{\bimK}
\safemath{\rvecl}{\bimL}
\safemath{\rvecm}{\bimM}
\safemath{\rvecn}{\bimN}
\safemath{\rveco}{\bomO}
\safemath{\rvecp}{\bimP}
\safemath{\rvecq}{\bimQ}
\safemath{\rvecr}{\bimR}
\safemath{\rvecs}{\bimS}
\safemath{\rvect}{\bimT}
\safemath{\rvecu}{\bimU}
\safemath{\rvecv}{\bimV}
\safemath{\rvecw}{\bimW}
\safemath{\rvecx}{\bimX}
\safemath{\rvecy}{\bimY}
\safemath{\rvecz}{\bimZ}

\safemath{\rvecxi}{\bmxi}
\safemath{\rveclambda}{\bmlambda}
\safemath{\rvecmu}{\bmmu}
\safemath{\rvectheta}{\bmtheta}
\safemath{\rvecphi}{\bmphi}

\safemath{\rmatA}{\bimA}
\safemath{\rmatB}{\bimB}
\safemath{\rmatC}{\bimC}
\safemath{\rmatD}{\bimD}
\safemath{\rmatE}{\bimE}
\safemath{\rmatF}{\bimF}
\safemath{\rmatG}{\bimG}
\safemath{\rmatH}{\bimH}
\safemath{\rmatI}{\bimI}
\safemath{\rmatJ}{\bimJ}
\safemath{\rmatK}{\bimK}
\safemath{\rmatL}{\bimL}
\safemath{\rmatM}{\bimM}
\safemath{\rmatN}{\bimN}
\safemath{\rmatO}{\bimO}
\safemath{\rmatP}{\bimP}
\safemath{\rmatQ}{\bimQ}
\safemath{\rmatR}{\bimR}
\safemath{\rmatS}{\bimS}
\safemath{\rmatT}{\bimT}
\safemath{\rmatU}{\bimU}
\safemath{\rmatV}{\bimV}
\safemath{\rmatW}{\bimW}
\safemath{\rmatX}{\bimX}
\safemath{\rmatY}{\bimY}
\safemath{\rmatZ}{\bimZ}

\safemath{\rmatDelta}{\bimDelta}
\safemath{\rmatLambda}{\bimLambda}
\safemath{\rmatPhi}{\bimPhi}
\safemath{\rmatSigma}{\bimSigma}
\safemath{\rmatOmega}{\bimOmega}
\safemath{\rmatTheta}{\bimTheta}

\usepackage{amssymb}
\usepackage{amsfonts}
\usepackage{mathrsfs}
\usepackage{xspace}
\usepackage{bm}
\usepackage{fancyref}
\usepackage{textcomp}

\usepackage{multirow}
\usepackage{stmaryrd}

\newenvironment{textbmatrix}{	\setlength{\arraycolsep}{2.5pt}%
								\big[\begin{matrix}}{\end{matrix}\big]%
								\raisebox{0.08ex}{\vphantom{M}}}

\def\be{\begin{equation}}
\def\ee{\end{equation}}
\def\een{\nonumber \end{equation}}
\def\mat{\begin{bmatrix}}
\def\emat{\end{bmatrix}}
\def\btm{\begin{textbmatrix}}
\def\etm{\end{textbmatrix}}

\def\ba#1\ea{\begin{align}#1\end{align}}
\def\bas#1\eas{\begin{align*}#1\end{align*}}
\def\bs#1\es{\begin{split}#1\end{split}} 
\def\bg#1\eg{\begin{gather}#1\end{gather}}
\def\bml#1\eml{\begin{multline}#1\end{multline}}
\def\bi#1\ei{\begin{itemize}#1\end{itemize}}

\DeclareMathOperator{\tr}{tr}				% trace
\DeclareMathOperator*{\argmin}{arg\;min}		% arg min
\newcommand{\tp}[1]{\ensuremath{#1^{T}}} 		% transpose
\newcommand{\herm}[1]{\ensuremath{#1^{H}}} 	% hermitian transpose
\newcommand{\inv}[1]{\ensuremath{#1^{-1}}} 	% inverse
\newcommand{\pinv}[1]{\ensuremath{#1^{\dagger}}} 	% Moore-Penrose pseudo-inverse
\safemath{\dirac}{\delta}					% Dirac delta
\safemath{\krond}{\dirac}					% Kronecker delta
\safemath{\upto}{\uparrow}
\safemath{\downto}{\downarrow}
\safemath{\iu}{j}							% imaginary unit
\safemath{\ev}{\lambda}						% eigenvalue
\safemath{\hilseqspace}{l^{2}}				% Hilbert sequence space
\newcommand{\banachfunspace}[1]{\setL^{#1}}	% Banach function space
\safemath{\hilfunspace}{\banachfunspace{2}}	% Hilbert function space
\safemath{\SNR}{\textsf{SNR}} 				% signal to noise ratio
\safemath{\PAR}{\textsf{PAR}} 				% signal to noise ratio
\safemath{\No}{N_0}							% noise spectral density
\safemath{\Es}{E_s}							% energy per symbol
\safemath{\Eb}{E_b}							% energy per bit
\safemath{\EbNo}{\frac{\Eb}{\No}}
\safemath{\EsNo}{\frac{\Es}{\No}}

\DeclareMathOperator{\CHop}{\ensuremath{\opH}} % channel operator
\safemath{\tvir}{\rndh_{\CHop}}				% time-varying impulse response
\safemath{\tvtf}{\rndl_{\CHop}}				% 	-''- transfer function
\safemath{\spf}{\rnds_{\CHop}}				% spreading function
\safemath{\bff}{H_{\CHop}}					% bi-freuqency function

\safemath{\ircf}{r_{h}}						% impulse response correlation fn.
\safemath{\tftvcf}{r_{s}}					% scattering function
\safemath{\tfcf}{r_{l}}						% time-frequency correlation fn.
\safemath{\bfcf}{r_{H}}						% bi-frequency correlation fn.

\safemath{\tcorr}{c_h}						% time-correlation function
\safemath{\scf}{c_{s}}						% spreading function
\safemath{\tfcorr}{c_{l}}					% transfer-function correlation
\safemath{\fcorr}{c_{H}}						% frequency-correlation function

\safemath{\mi}{I}							% mutual information
\safemath{\capacity}{C}						% capacity

\safemath{\normal}{\mathcal{N}}			% normal distribution
\safemath{\jpg}{\mathcal{CN}}			% jointly proper Gaussian
\safemath{\mchain}{\leftrightarrow}		% Markov chain
\safemath{\dB}{\,\mathrm{dB}}
\safemath{\dBm}{\,\mathrm{dBm}}
\safemath{\Hz}{\,\mathrm{Hz}}
\safemath{\kHz}{\,\mathrm{kHz}}
\safemath{\MHz}{\,\mathrm{MHz}}
\safemath{\GHz}{\,\mathrm{GHz}}
\safemath{\s}{\,\mathrm{s}}
\safemath{\ms}{\,\mathrm{ms}}
\safemath{\mus}{\,\mathrm{\text{\textmu}s}}
\safemath{\ns}{\,\mathrm{ns}}
\safemath{\ps}{\,\mathrm{ps}}
\safemath{\meter}{\,\mathrm{m}}
\safemath{\mm}{\,\mathrm{mm}}
\safemath{\cm}{\,\mathrm{cm}}
\safemath{\m}{\,\mathrm{m}}
\safemath{\W}{\,\mathrm{W}}
\safemath{\mW}{\, \mathrm{mW}}
\safemath{\J}{\,\mathrm{J}}
\safemath{\K}{\,\mathrm{K}}
\safemath{\bit}{\,\mathrm{bit}}
\safemath{\nat}{\,\mathrm{nat}}

\safemath{\define}{\triangleq}			% definition

\safemath{\equivalent}{\sim}
\safemath{\distas}{\sim}					% distributed according to
\safemath{\sdiff}{\Delta}				% symmetric set difference

\safemath{\reals}{\mathbb{R}}
\safemath{\positivereals}{\reals_{+}}
\safemath{\integers}{\mathbb{Z}}
\safemath{\posint}{\integers_{+}}
\safemath{\naturals}{\mathbb{N}}
\safemath{\posnaturals}{\naturals_{+}}
\safemath{\complexset}{\mathbb{C}}
\safemath{\rationals}{\mathbb{Q}}

\newcommand*{\fancyrefapplabelprefix}{app}		% Appendix
\newcommand*{\fancyrefthmlabelprefix}{thm}		% Theorem
\newcommand*{\fancyreflemlabelprefix}{lem}		% Lemma
\newcommand*{\fancyrefcorlabelprefix}{cor}		% Corollary
\newcommand*{\fancyrefdeflabelprefix}{def}		% Definition
\newcommand*{\fancyrefproplabelprefix}{prop}	% Proposition
\newcommand*{\fancyrefobslabelprefix}{obs}		% Observation 
\newcommand*{\fancyrefalglabelprefix}{alg}		% Algorithm
\newcommand*{\fancyrefasmlabelprefix}{asm}	    % Assumption
\newcommand*{\fancyreftbllabelprefix}{tbl}	    % Table

\frefformat{vario}{\fancyrefseclabelprefix}{Sec.~#1}
\frefformat{vario}{\fancyrefthmlabelprefix}{Thm.~#1}
\frefformat{vario}{\fancyreflemlabelprefix}{Lem.~#1}
\frefformat{vario}{\fancyrefcorlabelprefix}{Corr.~#1}
\frefformat{vario}{\fancyrefdeflabelprefix}{Def.~#1}
\frefformat{vario}{\fancyrefobslabelprefix}{Obs.~#1}
\frefformat{vario}{\fancyrefasmlabelprefix}{Ass.~#1}
\frefformat{vario}{\fancyreffiglabelprefix}{Fig.~#1}
\frefformat{vario}{\fancyrefapplabelprefix}{App.~#1} 
\frefformat{vario}{\fancyrefproplabelprefix}{Prop.~#1}
\frefformat{vario}{\fancyrefalglabelprefix}{Alg.~#1}
\frefformat{vario}{\fancyrefeqlabelprefix}{(#1)}
\frefformat{vario}{\fancyreftbllabelprefix}{Table~#1}

\safemath{\dictab}{[\,\dicta\,\,\dictb\,]}

\safemath{\ysig}{\bmy}
\safemath{\ysighat}{\hat{\ysig}}
\safemath{\ysigdim}{M}
\safemath{\xsig}{\bmx}
\safemath{\xsigdim}{N}
\safemath{\nx}{n_x}
\safemath{\zsig}{\bmz}
\safemath{\zsigdim}{\ysigdim}
\safemath{\rsig}{\bmr}
\safemath{\Adict}{\bA}
\safemath{\Adicttilde}{\widetilde{\Adict}}
\safemath{\Adictdim}{\outputdim\times\xsigdim}
\safemath{\avec}{\bma}
\safemath{\avectilde}{\tilde{\avec}}
\safemath{\Bdict}{\bB}
\safemath{\Bdicttilde}{\widetilde{\Bdict}}
\safemath{\Cdict}{\bC}
\safemath{\cvec}{\bmc}
\safemath{\Ddict}{\bD}
\safemath{\Ddictdim}{\ysigdim\times\xsigdim}
\safemath{\dvec}{\bmd}
\safemath{\Ddicttilde}{\widetilde{\bD}}
\safemath{\Bonb}{\bB}
\safemath{\bvec}{\bmb}
\safemath{\Bonbdim}{\ysigdim\times\ysigdim}
\safemath{\noise}{\bmn}
\safemath{\noisedim}{\ysigim}
\safemath{\err}{\bme}
\safemath{\errdim}{\ysigdim}
\safemath{\errset}{\setE}
\safemath{\nerr}{n_e}
\safemath{\delop}{\bP_\errset}
\safemath{\delopc}{\bP_{{\errset}^c}}

\safemath{\cplxi}{\imath}
\safemath{\cplxj}{\jmath}
\safemath{\dict}{\matD}
\safemath{\inputdim}{N}		% number of columns of dictionary D
\safemath{\outputdim}{M}		%number of rows of dictionary D
\safemath{\sparsity}{S}	%sparsity
\safemath{\inputdimA}{{N_a}}	%total number of elements in dictionary A
\safemath{\inputdimB}{{N_b}}	%total number of elements in dictionary B
\safemath{\elemA}{{n_a}}	%number of elements chosen from dictionary A
\safemath{\elemB}{{n_b}}	%number of elements chosen from dictionary B
\safemath{\resA}{\matR_a}	%restriction map to elements of dictionary A
\safemath{\resB}{\matR_b}	%restriction map to elements of dictionary B
\safemath{\subD}{\matS} %subdictionary
\safemath{\subA}{\matS_a} %subdictionary part of A
\safemath{\subB}{\matS_b} %subdictionary part of B
\safemath{\dicta}{\matA} 	% first subdictionary
\safemath{\dictb}{\matB} 	% second subdictionary
\safemath{\hollowS}{H}
\safemath{\hollowA}{H_a}
\safemath{\hollowB}{H_b}
\safemath{\cross}{Z}
\safemath{\coh}{\mu_d}			% coherence dictionary
\safemath{\coha}{\mu_a}			% coherence first subdictionary
\safemath{\cohb}{\mu_b}			% coherence second subdictionary
\safemath{\mubs}{\nu}	%block sub-coherence
\safemath{\cohm}{\mu_m} %mutual coherence
\safemath{\dictset}{\setD}	% set of dictionaries
\safemath{\dictsetp}{\dictset(\coh,\coha,\cohb)}	% set of dictionaries parametrized
\safemath{\dictsetgen}{\dictset_\text{gen}}
\safemath{\dictsetgenp}{\dictsetgen(\coh)}
\safemath{\dictsetonb}{\dictset_\text{onb}}
\safemath{\dictsetonbp}{\dictsetonb(\coh)}

\safemath{\leftside}{U}
\safemath{\rightsideA}{R_a}
\safemath{\rightsideB}{R_b}

\safemath{\indexS}{\setI_S} %set of indices participating in sub-dictionary S

\safemath{\nb}{n_b}			% cardinality of set of linearly independent columns of second dictionary
\safemath{\coeffa}{p_i}	%coefficients for columns of A
\safemath{\coeffb}{q_j}	%coefficients for columns of B
\safemath{\seta}{\setP}		% set of linearly independent columns of A
\safemath{\setb}{\setQ}     % set of linearly independent columns of B
\safemath{\setw}{\setW}	%set of n largest elements of w
\safemath{\setz}{\setZ}	%set of L-n largest elements of z
\safemath{\cola}{\veca}		% generic element of the dictionary A
\safemath{\colb}{\vecb}		% generic element of the dictionary B
\safemath{\cold}{\vecd}		% generic element of the dictionary D
\safemath{\inputvec}{\vecx} 	%coefficient vector (input)
\safemath{\error}{\vece}	%error vector
\safemath{\noiseout}{\vecz} 	%noisy output vector
\safemath{\inputvecel}{x}
\safemath{\inputveca}{\vecx_a}
\safemath{\inputvecb}{\vecx_b}
\safemath{\outputvec}{\vecy}	%output of Dictionary
\safemath{\lambdamin}{\lambda_{\mathrm{min}}}
\safemath{\elltwo}{\ell_2}
\safemath{\ellone}{\ell_1}
\safemath{\ellzero}{\ell_0}
\safemath{\ellinf}{\ell_\infty}
\safemath{\ellinftilde}{\ell_{\widetilde\infty}}
\safemath{\licard}{Z(\coh,\coha,\cohb)}
\safemath{\xsol}{\hat{x}}
\safemath{\xbord}{x_b}		%Solution at the border
\safemath{\xstat}{x_s}		%Solution stationary in l0 prob
\safemath{\xstatLone}{\tilde{x}_s}
\safemath{\order}{\mathcal{O}} %order notation (big O)
\safemath{\scales}{\Theta} %scales as
\safemath{\ones}{\mathbf{1}} %all ones matrix
\safemath{\zeroes}{\mathbf{0}} %all zeroes matrix
\safemath{\thlone}{\kappa(\coh,\cohb)} %treshold l1 problem
\safemath{\constoneA}{\delta} %constant in l1 theorem to save space
\safemath{\constoneB}{\epsilon} %constant in l1 theorem to save space
\safemath{\nlarge}{L}				   %num large elements
\safemath{\sumlarge}{S_\nlarge}
\safemath{\maxlarger}{P_\nlarge}	   % maximum in Gribonval and Nielsen
\safemath{\Pzero}{\textrm{P0}}	
\safemath{\Pone}{\textrm{P1}}
\safemath{\vecfir}{\vecw}			 % \vecv element of the kernel of the dictionary, \vecv=[\vecfir \vecsec]
\safemath{\vecsec}{\vecz}
\safemath{\elvecfir}{w}              % element of vecfir
\safemath{\elvecsec}{z}				 % element of vecsec
\safemath{\nlargefir}{n}
\safemath{\normout}{\gamma}
\safemath{\auxfun}{h}
\safemath{\supp}{\textrm{supp}}%support

\safemath{\indexa}{\ell}
\safemath{\indexb}{r}
\safemath{\indexc}{i}
\safemath{\indexd}{j}

\safemath{\project}{P}%projector

\IEEEoverridecommandlockouts

\safemath{\Herm}{\textnormal{H}}
\begin{document}

\bstctlcite{IEEEexample:BSTcontrol}

\title{
A Jammer-Resilient 2.87\,$\text{mm}^\text{2}$ 1.28\,MS/s 310\,mW Multi-Antenna Synchronization
ASIC in 65\,nm
}
\author{\IEEEauthorblockN{Flurin Arquint, Oscar Casta\~neda, Gian Marti, and Christoph Studer} \\[-0.3cm]
\thanks{This work has received funding from the Swiss State Secretariat for Education, Research, and Innovation (SERI) under the SwissChips initiative, and has also been supported by the European Commission within the context of the project 6G-REFERENCE (6G Hardware Enablers for Cell Free Coherent Communications and Sensing), funded under EU Horizon Europe Grant Agreement 101139155.
Contact author: O. Casta\~neda (e-mail: caoscar@ethz.ch)}
\IEEEauthorblockA{\emph{Department of Information Technology and Electrical Engineering, ETH Zurich, Switzerland}
} 
}

\maketitle

\begin{abstract}
We present the first ASIC implementation of jammer-resilient multi-antenna time synchronization. The ASIC implements a recent 
algorithm that mitigates jamming attacks on synchronization signals using multi-antenna processing.
Our design supports synchronization between a single-antenna transmitter and a 16-antenna receiver
while mitigating smart jammers with up to two transmit antennas. 
The fabricated 65\,nm ASIC has a core area of 2.87\,mm$^\text{2}$, consumes a power of 310\,mW, 
and supports a sampling rate of 1.28~mega-samples per second (MS/s). 

\end{abstract}

\section{Introduction}\label{sec:intro}

Our aspirations for a wireless future make jammer-resilient communications an imperative. 
Much attention has been paid to data transmission under jamming~\cite{do18a, hoang2021suppression, marti2023maed},
including the design of jammer-resilient data detectors as integrated circuits~\cite{bucheli2024vlsi}. 
However, to deploy these methods, the transmitter and receiver must be synchronized in time~\cite{schmidl97a}. 
This synchronization problem has received only little attention~\cite{bliss2009temporal,marti2024jass} 
and, to our knowledge, no corresponding integrated circuit has been developed so far.

\subsubsection*{Contributions}
We present an application-specific integrated circuit (ASIC) implementation of the recent JASS (short for Jammer-Aware SynchroniSation) algorithm~\cite{marti2024jass} 
for time synchronization between a single-antenna transmitter and a 16-antenna receiver under jamming. 
The jammer is mitigated by fitting an adaptive spatial filter to a time-windowed sequence of the receive signal. 
Our design can mitigate jammers with up to two antennas and, by using a secret synchronization sequence (consisting 
of 16 BPSK symbols), it is also able to mitigate smart jammers (i.e., jammers that 
rely on more sophisticated strategies than pure noise-like barrage jamming).
To our knowledge, our design is the first ASIC implementation of a jammer-resilient synchronization algorithm 
of \emph{any} kind. 

\section{Prerequisites} \label{sec:system}

\subsection{System Model}
We consider a multi-antenna receiver with a single-antenna transmitter (as in, e.g., a single-user uplink) 
under attack from a jammer that can potentially have multiple antennas. 
We model the receive signal at sample index $k=0,1,2,\dots$ as follows:
\begin{align}
	\bmy[k]	= \bmh s[k] + \bJ\bmw[k] + \bmn[k].
\end{align}
Here, $\bmy[k]\in\opC^B$ is the receive vector at the $B$-antenna receiver; $s[k]$ is the legitimate transmit signal, which corresponds to 
\begin{align}
	s[k] = \begin{cases}
 	0 &: k < L \\
 	\breve{s}_{k+1-L}	&: L \leq k < L+K \\
 	\text{undefined} &: L+K \leq k,
 \end{cases}
\end{align}
where $\breve\bms = \tp{[\breve{s}_1,\dots,\breve{s}_K]}\in\{\pm1\}^K$ is the length-$K$ synchronization sequence 
and $L\geq0$ represents the time at which the legitimate transmitter sends the start of the synchronization sequence; 
$\bmh\in\opC^B$ is the channel between the legitimate transmitter and the receiver;
$\bJ\in\opC^{B\times I}$ is the channel between the $I$-antenna jammer and the receiver; 
$\bmw[k]\in\opC^I$ is the jammer transmit vector; 
and $\bmn[k]\sim\setC\setN(\boldsymbol{0},\No\bI_B)$ is additive white Gaussian thermal noise with variance $\No$ per entry.

We assume that the synchronization sequence $\breve\bms$ is unknown {a priori} to the jammer 
(i.e.,~$\bmw[k]$ can only depend on $\breve{s}_{k'}$ if $k-L\geq k'$),
while $L$ is unknown {a priori} to the receiver and has to be estimated based on the receive vectors.
Specifically, the receiver has to solve the following problem: Based on a running receive sequence 
$\bmy[0],\dots,\bmy[\ell+K-1]$, the receiver must decide---for each~$\ell$---if $\ell$ is equal to $L$ or not. 
A \emph{false alarm} occurs when the receiver erroneously decides that $\ell=L$. 
A \emph{miss} occurs when the receiver erroneously decides that $\ell\neq L$. 
After the $(L+K-1)$th sample has been processed, the receiver has either successfully found 
$L$ or made an error (false alarm or miss), so there is no need to define $s[k]$ for $k\geq L+K$. 

\subsection{Jammer-Resilient Synchronization}
Our design is based on the JASS algorithm proposed recently in~\cite{marti2024jass}, which approximately solves the optimization problem
\begin{align}
	\argmin_{\ell\geq0} \ell \quad\text{such that}\quad \textnormal{max}_{\tilde\bP\in\setP_{I_{\textnormal{max}}}}\frac{\|\tilde\bP\bY_\ell\breve\bms^\ast\|_2^2}{\|\tilde\bP\bY_\ell\|_F^2}\geq \tau, \label{eq:opt_prob}
\end{align}
where $\bY_\ell=[\bmy[\ell],\dots,\bmy[\ell+K-1]]$ is the windowed receive signal,
$\tau\geq0$ is a detection threshold tunable for an optimal tradeoff between false alarms and misses,
$(\cdot)^\ast$~is the complex conjugate, 
and $\setP_{I_{\textnormal{max}}}=\{\bI_B - \bA\pinv{\bA}:\bA\in\opC^{B\times I_{\textnormal{max}}}\}$ is the set 
of orthogonal projections onto the $(B-I_{\textnormal{max}})$-dimensional subspaces of $\opC^B$, with $I_{\textnormal{max}}$ being the maximum number of 
jammer antennas that can be mitigated ($I\leq I_{\textnormal{max}}<B$); $\bI_B$, the \mbox{$B\!\times\! B$}~identity matrix; and $\pinv{(\cdot)}$, the Moore-Penrose~inverse.

To approximately solve \eqref{eq:opt_prob}, JASS performs the following operations for $\ell=0,1,\dots,\ell_{\textnormal{max}}$: 
First, it estimates the interference subspace by computing the~$I_{\textnormal{max}}$ principal eigenvectors of~\mbox{$\biLambda=\|\breve\bms\|_2^2\bY_\ell\herm{\bY_\ell}-\bY_\ell\breve\bms^\ast\tp{\breve\bms}\herm{\bY_\ell}$}.
These principal eigenvectors are collected into a matrix~$\bA$, so that the estimated interference subspace can be removed through the projection matrix~\mbox{$\tilde\bP=\bI_B-\bA\pinv{\bA}\in\setP_{I_{\textnormal{max}}}$}.
Then, JASS computes the {score} 
$\|\tilde\bP\bY_\ell\breve\bms^\ast\|_2^2/\|\tilde\bP\bY_\ell\|_F^2$. 
If the score reaches the threshold~$\tau$, then JASS terminates and declares $\ell$ to be the delay index $L$ at which the start of the synchronization sequence~$\breve\bms$ was received. If the threshold~$\tau$ is not reached after evaluating $\ell_{\textnormal{max}}$ candidate indexes, then JASS terminates and declares a miss.
For a proper motivation of JASS, its algorithmic details, and corresponding success guarantees, see~\cite{marti2024jass}.

To enable an efficient hardware implementation, the JASS algorithm from~\cite{marti2024jass} was reorganized into \fref{alg:jass}.
Instead of explicitly computing $\biPhi=\bY_\ell\herm{\bY_\ell}$ for each $\ell$, $\biPhi$ is iteratively updated (line 17).
Each of the $I_{\textnormal{max}}$ principal eigenvectors of $\biLambda$ is computed using $t_{\textnormal{max}}$ iterations of the power method (lines~5 to~10).
Our ASIC is dimensioned to mitigate jammers with up to $I_{\textnormal{max}}=2$ antennas. This parameter choice simplifies the computation of the matrix inverse $\inv{(\herm{\bA}\bA)}=(1-|\tilde{b}|^2)^{-1}\tilde{\bB}$ (line 11), which is needed to compute $\pinv{\bA}=\inv{(\herm{\bA}\bA)}\herm{\bA}$.
To avoid the division required to compute the score, we instead take the numerator $N$ (line 13) and denominator $D$ (line 14) of the score and check whether $N-D\tau\geq 0$ (line 15) instead of $N/D\geq \tau$.
Note that in these computations, the projection matrix~\mbox{$\tilde\bP=\bI_B-\bA\pinv{\bA}$} is never computed explicitly.

\begin{algorithm}[tp]
	\caption{Jammer-Aware Synchronisation (JASS)}
	\label{alg:jass}
	\begin{algorithmic}[1]
	\State $\biPhi = \bY_0\herm{\bY_0}$
	\For{$\ell=0,1,2,\dots,\ell_{\textnormal{max}}$}
	\State $\bmc_\ell = \bY_\ell\,\breve\bms^\ast$
	\State $\biLambda = \|\breve\bms\|_2^2\biPhi - \bmc_\ell\herm{\bmc_\ell}$
	\For{$i=1,\dots,I_{\textnormal{max}}(=2)$}
         	\State $\bma_i \leftarrow \text{PRNG}$
	\For{$t=1,\dots,t_{\textnormal{max}}(=2)$}
		\State $\bma_i' = \biLambda\bma_i$
		\State $\bma_i = \bma_i'/\|\bma_i'\|_2$
	\EndFor
		\State $\biLambda \leftarrow \biLambda - \bma_i'\herm{\bma_i}$
	\EndFor
	\vspace{-0.1mm}
	\State $\bA = [\bma_1,\bma_2]$; ~$\tilde{b} = \herm{\bma_1}\bma_2$; ~$\tilde\bB =$\scriptsize$\begin{bmatrix}1 & -\tilde{b}\\ -\tilde{b}^\ast & 1\end{bmatrix}$\normalsize
	\vspace{-0.1mm}
	\State $\bmv = \herm{\bA}\bmc_\ell$; ~$\bW = \herm{\bA}\biPhi$ 
	\State $N = (1-|\tilde{b}|^2)\|\bmc_\ell\|_2^2-\herm{\bmv}\tilde\bB\bmv$
	\State $D = (1-|\tilde{b}|^2)\tr(\biPhi) - \tr(\widetilde\bB\bW\bA)$
        \If{$N-D\tau\geq0$}\label{alg:jass_score}
        \vspace{0.1mm}
            \State \textbf{return} $\ell$
        \EndIf
        \State $\biPhi \leftarrow \biPhi - \bmy[\ell]\herm{\bmy[\ell]} + \bmy[\ell+K]\herm{\bmy[\ell+K]}$
	\EndFor 
	\end{algorithmic}
\end{algorithm}

\section{VLSI architecture} \label{sec:arch}
Our ASIC is designed to execute \fref{alg:jass} for a receiver with $B=16$ antennas, using a programmable synchronization sequence consisting of $K=16$ BPSK symbols under the interference of a jammer with at most $I_{\textnormal{max}}=2$ antennas. The detection threshold $\tau$, as well as the maximum number $\ell_{\textnormal{max}}$ of candidate indexes to be evaluated, is programmable. The number of iterations for the power method is fixed to $t_{\textnormal{max}}=2$, since it delivers practically the same performance as an exact eigenvalue decomposition, as used in the original algorithm~\cite{marti2024jass}.

\begin{figure}[t]
	\includegraphics[width=0.999\linewidth]{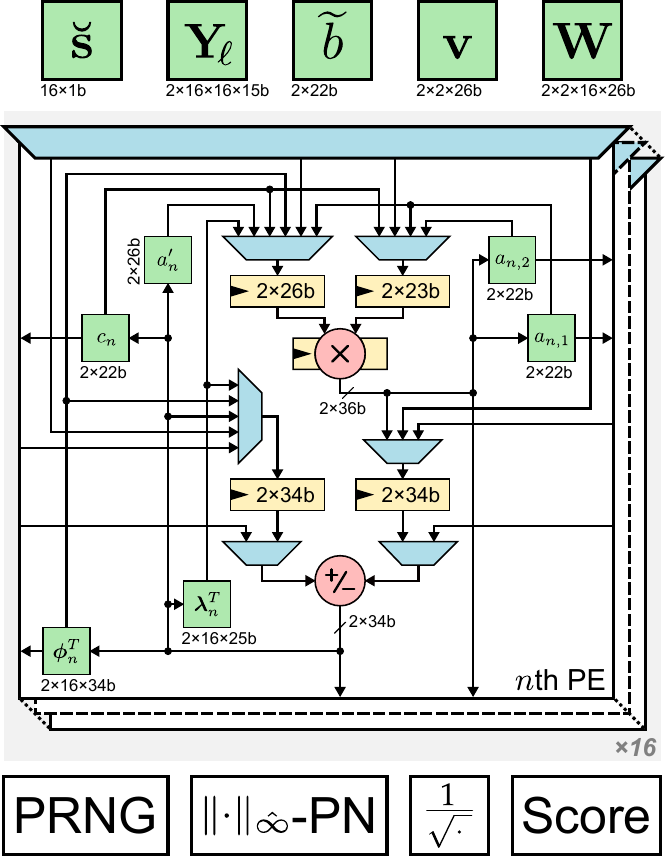}
	\caption{Overview of the JASS architecture, consisting of 16 reconfigurable processing elements (PEs), a pseudorandom number generator (PRNG), a $\|{\cdot}\|_{\hat{\infty}}$--pseudonormalization~(PN) module, an inverse square root module, and the score module. Blocks in green correspond to flip-flops (FFs) and FF arrays. The $\tp{\bmlambda}_{n}$ and $\tp{\bmphi}_{n}$ FF arrays store the $n$th row of $\biLambda$ and $\biPhi$, respectively.}
	\label{fig:arch_overview}
\end{figure}

\fref{fig:arch_overview} provides an overview of the top-level hardware architecture, which consists primarily of $16$ reconfigurable processing elements~(PEs). Each PE includes a pipelined complex-valued multiplier, a complex-valued adder, and flip-flop~(FF) arrays for local storage. The number of PEs was chosen to match the largest matrix dimension (in our case, \mbox{$B=K=16$}) in order to reduce latency. Besides the PEs, the architecture contains a module for complex-valued pseudorandom number generation (PRNG; cf. \fref{sec:prng_module}), a $\|{\cdot}\|_{\hat{\infty}}$--pseudonormalization~(PN) module (cf. \fref{sec:pn_module}), an inverse square root module (cf. \fref{sec:inv_sqrt_module}), and a score module (cf. \fref{sec:operation}). 
Finally, the architecture integrates FF arrays to store $\breve{\bms}$, $\bY_\ell$, $\tilde{b}$, $\bmv$, and $\bW$. Every PE can access the contents of these FF arrays, as well as the outputs from other PEs, through an interconnect network (represented by the large multiplexers at the top of each PE in \fref{fig:arch_overview}).

\begin{figure}[t]
	\hspace{-0.25cm}
	\begin{minipage}{0.475\linewidth}
    	\subfigure[Accumulator configuration of the PEs for the operation $\biLambda \bma$.]{
        	\includegraphics[width=\textwidth]{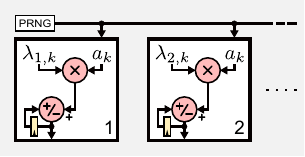}
        	\label{fig:acc_mode}
    	}
    \end{minipage}
	\hspace{0.2cm}
    \begin{minipage}{0.475\linewidth}
    	\subfigure[PEs configured to multipy-and-subtract for the operation $\biLambda - \bma'\herm{\bma}$.]{
        	\includegraphics[width=1\linewidth]{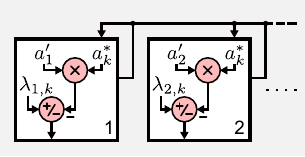}
        	\label{fig:mult_sub_mode}
    	}
    \end{minipage}
	
	\begin{minipage}{\linewidth}
		\hspace{-0.2075cm} 
    	\subfigure[Adder tree configuration of the PEs with extra pipeline registers.]{
        	\includegraphics[width=\linewidth]{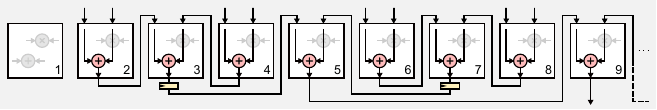}
        	\label{fig:tree_mode}
    	}
    \end{minipage}
	\caption{Illustration of three different PE configurations for the execution of the operations of the JASS algorithm.}
    \label{fig:jass_arch_modes}
\end{figure}

\subsection{Operation}\label{sec:operation}
Most operations of \fref{alg:jass} boil down to matrix-vector products (lines 3, 8, 12, 13, and 14)
and rank-one matrix updates of the form $\biLambda - \bma'\herm{\bma}$ (lines 4, 10, and 17).
For the matrix-vector product~$\biLambda\bma$ on line 8 (which is executed four times per delay index $\ell$), the PEs are configured as depicted in \fref{fig:acc_mode}:
In each clock cycle $k$, a new column $\bmlambda_k$ of $\biLambda$ is scaled by $a_k$ and accumulated to the previous result, until $\biLambda\bma=\sum^{16}_{k=1} \bmlambda_{k} a_k$ is completed. Due to pipelining registers, computing $\biLambda\bma$ takes 19 clock cycles.
A rank-one matrix update $\biLambda\leftarrow\biLambda - \bma'\herm{\bma}$ also takes 19 clock cycles and is performed by calculating, in clock cycle~$k$, the column $\bmlambda_{k}\leftarrow\bmlambda_{k} -\bma'a_{k}^\ast$ by using the $16$~PEs in parallel as visualized in \fref{fig:mult_sub_mode}.

The computation of $\bmc_\ell$ on line 3 follows the configuration in \fref{fig:acc_mode}, but the signs of the columns are simply adjusted instead of using the multipliers, since $\breve\bms$ is a BPSK sequence.
For the same reason, the matrix rescaling $\|\breve\bms\|_2^2\biPhi$ on line~4 amounts to a left shift by $\log_2 16=4$ bits. 
The PRNG operation from line 6 is described in \fref{sec:prng_module}.
The vector normalization from line 9 is performed in two steps: The first step consists of a pseudonormalization 
(described in \fref{sec:pn_module}) to reduce the bitwidth of the entries of $\bma_i'$ for the subsequent exact normalization. 
The exact normalization is performed by multiplying the pseudonormalized $\bma_i'$ with the inverse square root (see \fref{sec:inv_sqrt_module}) of its squared $\ell_2$-norm $\|\bma_i'\|_2^2$, with $\|\bma_i'\|_2^2$ being computed using the adder tree configuration shown in \fref{fig:tree_mode}.
The adder tree configuration is also used to compute the inner products on lines 11 and~13, the matrix-vector and matrix-matrix products on line~12, and the trace operations $\tr(\cdot)$ on line~14.
Due to pipelining, a single inner product has a latency of 5 clock cycles.
Finally, the score module utilizes its own real-valued multipliers and subtractors to complete the computation of $N$, $D$, and $N-D\tau$ (lines 13--15), to determine if the programmable threshold $\tau$ was reached.
In total, 268 clock cycles are required to process one delay index $\ell$, i.e., to evaluate whether or not the synchronization sequence occurred within the windowed receive signal $\bY_\ell$.

\subsection{Pseudorandom Number Generator~(PRNG)}\label{sec:prng_module}
The PRNG consists of two 32-bit xorshift~\cite{marsaglia2003xorshift} blocks as visible in \fref{fig:jass_arch_prng} to produce both real and imaginary parts of the output within one clock cycle. For the first xorshift block, the initial state for the first delay index $\ell$ can be programmed, while for all subsequent indices, the output state from the second xorshift is fed back to the input of the first xorshift.

\subsection{$\|{\cdot}\|_{\hat{\infty}}$--Pseudonormalization~(PN) Module}\label{sec:pn_module}
The jammer's large dynamic range requires the PEs' datapath to have larger bitwidths \cite{marti2024fundamental}. 
This issue is particularly exacerbated when computing quantities directly associated to the jammer's power, such as $\|\bma_i'\|^2_2$. Instead of using additional PEs with even larger bitwidths to compute $\|\bma_i'\|^2_2$ as it was done in~\cite{bucheli2024vlsi}, we use the $\|{\cdot}\|_{\hat{\infty}}$--PN module to scale the entries of the vector $\bma_i'$ down into a known range, thereby reducing the bitwidth of $\|\bma_i'\|^2_2$.
To arrive at a low-complexity, hardware-friendly normalization,
we use the pseudonorm $\|\cdot\|_{\hat\infty}\triangleq 2^n$, where ${n=\lfloor \log_2(\textnormal{max}\{\|\Re\{\cdot\}\|_{\infty}, \|\Im\{\cdot\}\|_{\infty}\}) \rfloor}$ and $\|\cdot\|_{\infty}$ is the infinity norm.
As illustrated in \fref{fig:jass_arch_pnorm}, normalization with respect to this pseudonorm amounts to an arithmetic right shift by 
$n$ bits, where $n$ is simply computed 
using an OR-tree to combine the absolute values of the real and imaginary parts of all entries of $\bma_i'$, 
followed by a leading-one detector (LOD) with base 2 (LOD2).
After the $\|{\cdot}\|_{\hat{\infty}}$--PN module, the real and imaginary parts of the entries of $\bma_i'$ are in the range $[-2,2)$  and are represented with 21 bits instead of 34 bits.

\subsection{Inverse Square Root Module}\label{sec:inv_sqrt_module}
\begin{figure}[t]
	\hspace{-0.25cm}
    \begin{minipage}{0.495\linewidth}
    	\subfigure[PRNG producing complex-valued numbers using xorshift.]{
        	\includegraphics[width=1.0\linewidth]{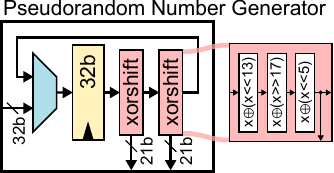}
        	\label{fig:jass_arch_prng}
    	}\vspace{0.22cm}
    	\subfigure[$\|{\cdot}\|_{\hat{\infty}}$--pseudonormalization module to rescale complex-valued vectors.]{
       		\includegraphics[width=1.0\linewidth]{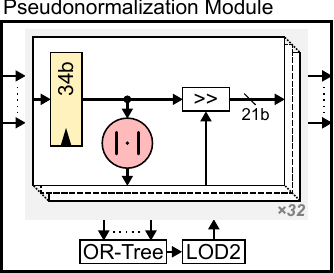}
        	\label{fig:jass_arch_pnorm}
    	}
    \end{minipage}
    \hspace{0.24cm}
    \begin{minipage}{0.45\linewidth} 
    	\subfigure[Module for computing the inverse square root using a LUT and a Newton-Raphson iteration.]{
        \includegraphics[width=1.0\linewidth]{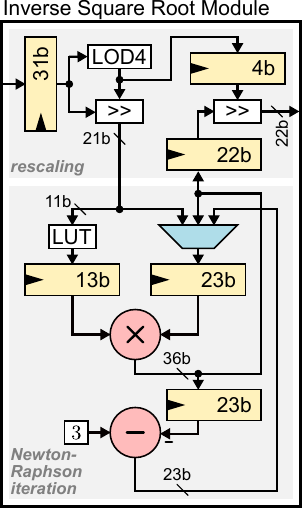}
        \label{fig:jass_arch_inv_sqrt}
    }
    \end{minipage}
    \caption{Internal structure of the (a) PRNG module, (b) $\|{\cdot}\|_{\hat{\infty}}$--PN module, and (c) inverse square root module of the JASS hardware architecture.}
    \label{fig:jass_arch_modules}
\end{figure}

The inverse square root module is shown in \fref{fig:jass_arch_inv_sqrt}. First, the input $x$ is rescaled as $x'=x/2^{2\alpha}$, where $\alpha\in\mathbb{Z}$ is found using a LOD with base~4 (LOD4), so that $x'\in [0.25, 1)$. Then, an initial estimate of $1/\sqrt{x'}$ is fetched from a look-up table~(LUT) and refined with one Newton-Raphson iteration: $y = y_{\text{LUT}}(3-y_{\text{LUT}}^2x')/2$. Here, a dedicated real-valued multiplier and subtractor are used to avoid the less efficient complex-valued datapath of the PEs. %Finally, the result $y=1/\sqrt{x}$ is scaled back using a base-2 shift according to the LOD4's output.
Finally, $y\approx1/\sqrt{x'}=2^\alpha/\sqrt{x}$ is scaled back to obtain the desired $1/\sqrt{x}$.

\section{ASIC Implementation Results}\label{sec:impl}
\subsection{Synchronization Error Rate~(SER) Performance}
\begin{figure}[tp]
	\hspace{-0.25cm}
    \subfigure[Delayed-spoofing, $\rho = 0$\,dB]{
        \includegraphics[width=0.485\linewidth]{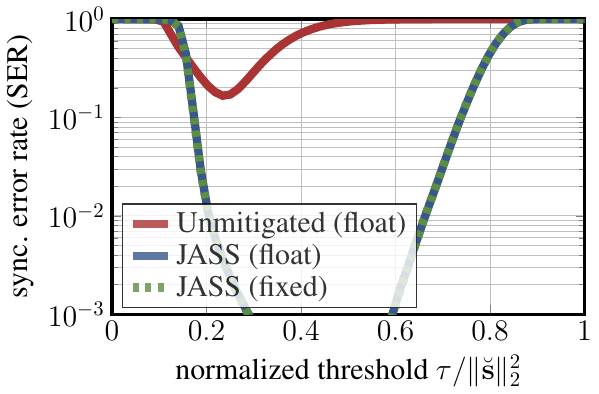}
        \label{fig:delayed_spoofer_rhodB_0}
    }
    \hspace{-0.22cm}
    \subfigure[Antenna-switch., $\rho = 10$\,dB]{
        \includegraphics[width=0.485\linewidth]{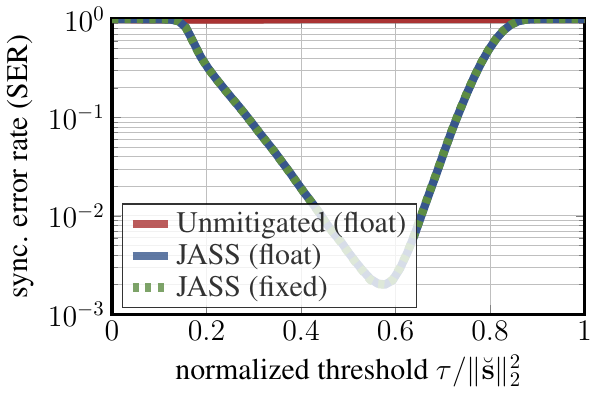}
        \label{fig:antenna_switching_rhodB_10}
    }

    \hspace{-0.25cm}
    \subfigure[Erratic, $\rho = 20$\,dB]{
        \includegraphics[width=0.485\linewidth]{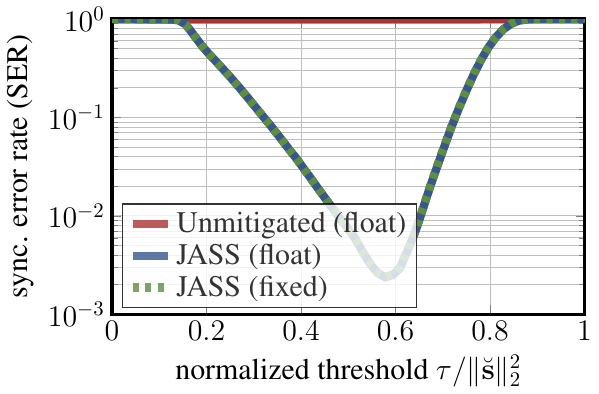}
        \label{fig:erratic_rhodB_20}
    }
    \hspace{-0.22cm}
    \subfigure[Barrage, $\rho = 30$\,dB]{
        \includegraphics[width=0.485\linewidth]{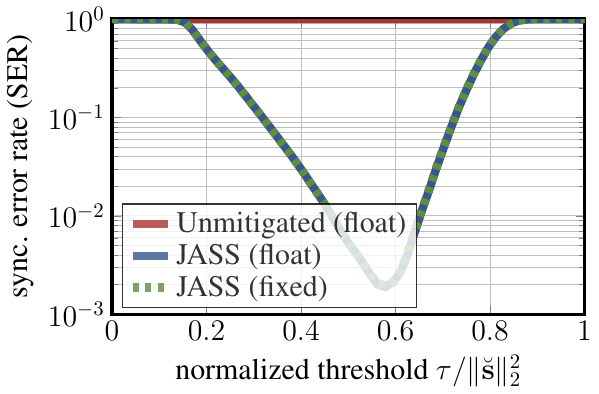}
        \label{fig:barrage_rhodB_30}
    }
	\caption{Synchronization error rate (SER) performance against a two-antenna jammer at a signal-to-noise ratio (SNR) of $5$\,dB, 
	for different jammer types and jammer-to-signal ratios~$\rho$.
	The delayed-spoofing jammer repeats the synchronization sequence $\breve\bms$ with one sample delay; %\protect\footnote{}
	the antenna-switching jammer transmits Gaussian symbols using sometimes one antenna, sometimes the other; 
	the erratic jammer transmits Gaussian symbols at random times and is otherwise silent; 
	and the barrage jammer transmits white Gaussian noise. 
	}
    \label{fig:ter_plots}
\end{figure}

\fref{fig:ter_plots} shows the synchronization error rate (SER) as a function of the threshold\footnote{A large threshold $\tau$ entails a low probability of a false alarm but a high probability of a miss, while the opposite holds for a small $\tau$---the optimal $\tau$ balances between false alarms and misses.} $\tau$ for our fixed-point hardware implementation of JASS under different types of two-antenna jammers with different jammer-to-signal ratios~$\rho$. 
In all four jammer scenarios, JASS consistently and substantially outperforms an unmitigated synchronization approach (i.e., setting the matrix $\bA$ to the all-zeroes matrix), which fails due to jamming. It is also evident from \fref{fig:ter_plots} that the fixed-point arithmetic\footnote{The bitwidths of the fixed-point implementation are indicated in Figs.~\ref{fig:arch_overview} and \ref{fig:jass_arch_modules}, where the $(2{\times})$ accounts for real and imaginary components.} implemented in the JASS ASIC incurs in virtually no performance loss compared to a floating-point baseline.

\begin{figure}[tp]
	\begin{minipage}[tp]{0.49\linewidth}
    	\includegraphics[width=1\linewidth]{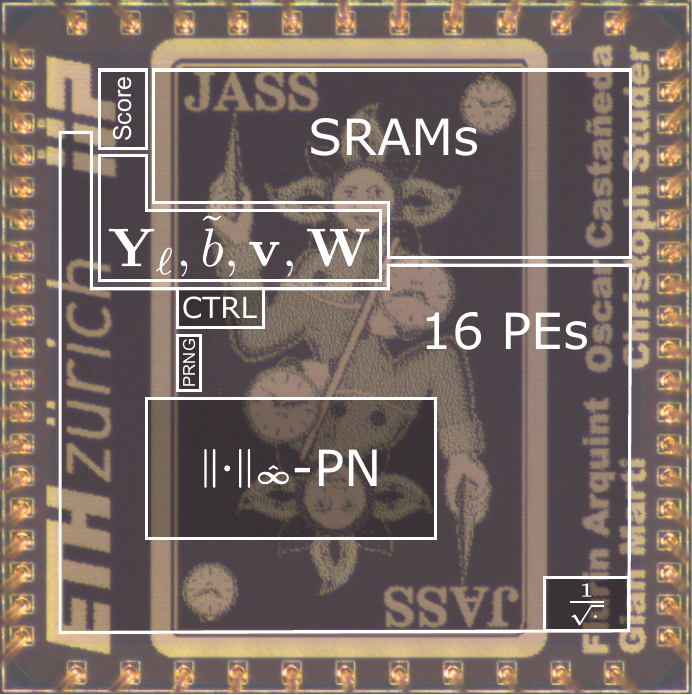}
    	\vspace{-6mm}
		\caption{Micrograph of the $2\,\text{mm}\times2\,\text{mm}$ JASS ASIC in TSMC 65\,nm LP with highlighted modules.}
		\label{fig:jass_chip_micrograph}
    \end{minipage}
	\hspace{0.2cm}
    \begin{minipage}[tp]{0.46\linewidth}
    	\includegraphics[width=1\linewidth]{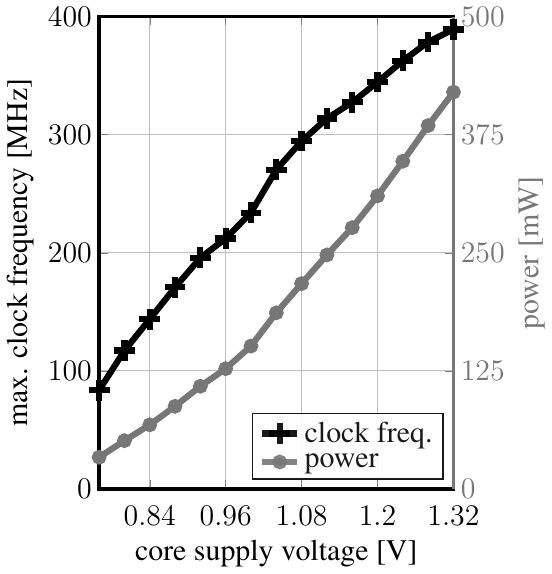}
    	\vspace{-6mm}
		\caption{Measured maximum clock frequency and power consumption versus core supply voltage.}
		\label{fig:supply_vs_freq_and_power}
    \end{minipage}
\end{figure}

\subsection{ASIC Measurements}
\fref{fig:jass_chip_micrograph} presents the micrograph of the $4\,\text{mm}^2$ JASS ASIC fabricated in TSMC 65\,nm LP, with its different modules, including the control unit and SRAMs, highlighted within the $2.87\,\text{mm}^2$ core area.
The SRAMs are used to store up to 1024 receive samples ${\bmy[k]}$. 
 At nominal core supply voltage of $1.2\,\text{V}$ and a room temperature of $300$\,K, the ASIC achieves a maximum clock frequency of $344$\,MHz, which corresponds to a sampling rate of $1.28$ mega-samples per second (MS/s). At this operating point, the ASIC (including SRAMs) consumes $310\,\text{mW}$  when facing a two-antenna barrage jammer with \mbox{$\rho=30$\,dB}. 
\fref{tbl:performance_figures} summarizes the performance metrics of the JASS ASIC, while Fig.~\ref{fig:supply_vs_freq_and_power} shows how the clock frequency and power scale with the core supply voltage.

As JASS is the first jammer-resilient synchronization ASIC, a comparison with other designs is not possible. Nevertheless, we contextualize our results by noting that, after technology normalization, our JASS ASIC (including SRAMs) occupies 9\% of the area of the jammer-resilient MIMO detector from~\cite{bucheli2024vlsi}, while consuming 8\% of its power when operating at the same clock frequency. 
We emphasize that jammer resilience comes at the cost of a significantly lower hardware efficiency and throughput~\cite{bucheli2024vlsi}. This remains true for our JASS ASIC, especially when considering that it is able to counteract smart jammers with more than one antenna. 
Still, higher sampling rates could be achieved at the cost of silicon area by having several instances of the proposed JASS architecture operating in parallel on different delay indexes $\ell$, or simply by reimplementing JASS in a more advanced technology node.

\begin{table}[tp]
\centering
\caption{Measurement results for the JASS ASIC}
\vspace{-0.2cm}
\begin{tabular}{@{}p{0.27\linewidth}c|p{0.29\linewidth}c@{}}
	\toprule
	Receiver antennas $B$ & $16$ & Core area [$\text{mm}^2$] &$2.87$\\
	Sync. seq. length $K$ & $16$ & Frequency [MHz] & $344$ \\
	Jammer antennas $I$ &$0$--$2$ & Sampling rate [MS/s] & $1.28$ \\
	Technology [nm] &$65$ & Power [mW] & $310$ \\
	Supply [V] &1.2 & Area eff. [MS/s/mm$^\text{2}$] & $0.45$\\
	& & Energy/sample [nJ/S] & $242$\\
\bottomrule
\end{tabular}
\label{tbl:performance_figures}
\end{table}

\vfill

\section{Conclusions}
We have presented the \emph{first} ASIC for jammer-resilient time synchronization reported in the open literature. Even when confronted with smart and strong jammers, our JASS ASIC performs accurate time synchronization at $1.28$\,MS/s when implemented on $2.87\,\text{mm}^2$ in a 65\,nm technology node.

\end{document}